\documentclass[preprint,showpacs,amsmath,amssymb,nofootinbib]{revtex4}
\usepackage{changes}
\usepackage{amsfonts}
\usepackage{amsmath,graphicx,color,epsfig}

\newcommand{\be}{\begin{equation}}
\newcommand{\ee}{\end{equation}}
\newcommand{\bea}{\begin{eqnarray}}
\newcommand{\eea}{\end{eqnarray}}
\newcommand{\ba}{\begin{array}}
\newcommand{\ea}{\end{array}}
\newcommand{\bi}{\begin{itemize}}
\newcommand{\ei}{\end{itemize}}

\newcommand{\mca}{{\mathcal A}}

\newcommand{\mcf}{{\mathcal F}}
\newcommand{\mci}{{\mathcal I}}

\newcommand{\mcn}{{\mathcal N}}

\renewcommand{\vec}[1]{\mbox{\boldmath $#1 \!\!$ \unboldmath}}

\newcommand{\g}{\gamma}
\newcommand{\nslash}{\kern 0.2 em n\kern -0.50em /}
\newcommand{\kslash}{\kern 0.2 em k\kern -0.45em /}
\newcommand{\qslash}{\kern 0.2 em q\kern -0.45em /}
\newcommand{\pslash}{\kern 0.2 em p\kern -0.50em /}
\newcommand{\rslash}{\kern 0.2 em r\kern -0.50em /}
\newcommand{\sslash}{\kern 0.2 em s\kern -0.50em /}
\newcommand{\Sslash}{\kern 0.2 em S\kern -0.50em /}
\newcommand{\Pslash}{\kern 0.2 em P\kern -0.50em /}
\newcommand{\Dslash}{\kern 0.2 em D\kern -0.65em /\kern 0.15em}
\newcommand{\lf}{\left}
\newcommand{\rg}{\right}

\newcommand{\slim}{\mskip 1.5mu}              

\newcommand{\cdott}{{\mskip -1.5mu} \cdot {\mskip -1.5mu}}
\begin{document}
\title{Parton distributions and $\cos 2\phi_h$ asymmetry \\ induced by anomalous photon-quark coupling }

\author{Xu Cao \footnote{caoxu@impcas.ac.cn}}
\affiliation{CAS Key Laboratory of High Precision Nuclear Spectroscopy and Center for Nuclear Matter Science, Institute of Modern Physics, Chinese Academy of Sciences, Lanzhou 730000, China }
\affiliation{
State Key Laboratory of Theoretical Physics, Institute of Theoretical Physics, Chinese Academy of Sciences, Beijing 100190, China}

\begin{abstract}

  Abstract: In the spectator models of the nucleon with scalar and axial-vector diquarks, we show that the effect of Pauli coupling in photon-quark vertex to the parton distribution functions (PDFs) of nucleon and azimuthal asymmetry in the unpolarized semi-inclusive deep inelastic scattering (SIDIS). This anomalous coupling gives obvious contribution to unpolarized and polarized PDFs, and also leads to a $\cos 2\phi_h$ azimuthal asymmetry proportional to the squared Pauli form factor, due to the helicity flip of the struck quark. After determining the model parameters by fitting PDFs to the global fits, this new distribution for $\cos 2\phi_h$ asymmetry is given numerically. In the framework of transverse momentum dependence (TMD), we find that it is positive and of a few percent in the kinematical regime of HERMES and COMPASS collaborations, in the same order of magnitude with Cahn effect.

\end{abstract}
\maketitle

\section{Introduction} \label{sec:intro}

The semi-inclusive deep inelastic scattering (SIDIS) not only plays an essential role in studying the parton distribution functions (PDF) of nucleon, but also its azimuthal asymmetries are important observables related to the transverse spin of quarks inside target hadrons~\cite{Barone:2010zz}. Two azimuthal moments are present in the unpolarized  differential cross sections, with the angular dependence of $\cos \phi_h$ and $\cos 2\phi_h$, respectively (here and afterwards $\phi_h$ is the azimuthal angle of produced hadron in the scattering plane). Two main mechanisms for the origin of these asymmetries are usually considered when QCD factorization is applicable. One of them involves the convolution of Boer-Mulders function~\cite{Boer:1997nt}, which measures the transverse polarization asymmetry of quarks inside an unpolarized hadron, and the spin-dependent Collins fragmentation function (FF) of the produced hadron~\cite{Collins:1992kk,Collins:2002kn}. The other one is in terms of Cahn effect, related to the non-collinear transverse-momentum kinematics. While the former effect is the leading twist one, the latter is kinematically of higher twist. These functions should be either calculated within some phenomenological models~\cite{Bacchetta:2003rz,Bacchetta:2008af,Barone:2005kt,Gamberg:2007wm} or be extracted directly from experiments~\cite{Anselmino:2005nn,Anselmino:2013lza,Barone:2008tn,Barone:2009hw,Barone:2015ksa,Zhang:2008ez,Zhang:2008nu,Lu:2009ip,Echevarria:2014xaa,Boglione:2011wm,Pasquini:2011tk,Schweitzer:2010tt,Wang:2017onm}.

Dirac vector coupling is widely explored in the quark electromagnetic and chromomagnetic current, but the tensor coupling with respect to anomalous Pauli form factor is less recognized. These Pauli couplings can flip the quark helicity, which is one of the important ingredients to induce single-spin asymmetries (SSA) in SIDIS. It was found that the Pauli couping in quark-gluon vertex from various sources leads to a large SSA in quark-quark scattering~\cite{Kochelev:2013zoa}. In the hadronic level, possible role of the Pauli-type soft quark-gluon interaction on SSAs in SIDIS was investigated in spectator model~\cite{Hoyer:2005ev}. Within the same model adopting the scalar and axial-vector diquark models for the nucleon, it was recently demonstrated that Pauli couplings in both quark-photon and quark-gluon vertices produce considerable SSA in SIDIS, whose azimuthal dependencies are the same as that usually called the Collins and Sivers effects~\cite{Cao:2017bdi}. Alternatively approaches, e.g. instanton background field~\cite{Ostrovsky:2004pd}, MIT bag model~\cite{Cherednikov:2006zn} and single instanton approximation~\cite{Qian:2015wyq}, arrive at similar conclusions. This remarkable mechanism would be promising for our understanding of the large SSAs observed in high energy hadronic reactions and in SIDIS~\cite{Barone:2010zz}.

The Pauli couplings can be originated from instantons, a small size strong gluonic fluctuation in the QCD vacuum~\cite{Schafer:1996wv,Diakonov:2002fq}. These non-trivial topological structure of QCD equations generate a very large quark chromomagnatic moment, directly resulting into the Pauli coupling in quark-gluon vertex~\cite{Kochelev:1996pv}. It was revealed that this anomalous interaction would play important role in gluon distributions in the nucleon~\cite{Kochelev:2015pqd}, in quark-gluon plasma~\cite{Kochelev:2015jba}, in high energy elastic $pp$ scattering~\cite{Kochelev:2013csa} and $pp \to \pi^0 X$ reaction~\cite{Kochelev:2015pha}. In these analyses it was suggested that this quark-gluon interaction would cause the explicit breakdown of transverse-momentum-dependent (TMD) factorization. In addition, the instantons would also give rise to the Pauli coupling in quark--photon vertex, as illustrated within the nonperturbative approach~\cite{Kochelev:2003cp,Zhang:2017zpi}. The instanton liquid model~\cite{Schafer:1996wv,Diakonov:2002fq} is widely implemented in the qualitative calculations.

The purpose of this paper is to consider what specifical effects the Pauli coupling of quark-photon interaction may have on PDFs and unpolarized SIDIS. The contribution of this interaction are expected to be complementary to the Dirac couplings. In Sec.~\ref{sec:resul} we give our analytical results of TMD PDFs in spectator models and try to extract the model parameters from the global PDF fits. In Sec.~\ref{sec:numer}, we give our new $\cos 2\phi_h$ asymmetry in the framework of TMD factorization and compare it to the Cahn effect. We conclude briefly in Sec.~\ref{sec:conclusion}.

\section{TMD Parton Distributions in spectator models} \label{sec:resul}

\begin{figure}
\begin{center}
{\includegraphics*[width=6.cm]{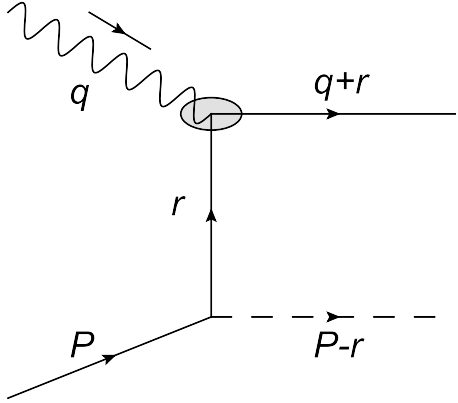}}
\caption{ Diagrammatic representation of the DIS process in our calculation. The blob represents the interaction vertices with both Dirac and Pauli couplings.
\label{fig:SSAFeynman}}
\end{center}
\end{figure}

In this work we employ the parton model of nucleon by using the spectator model with scalar and axial-vector diquarks~\cite{Jakob:1997wg}, as depicted in Fig~\ref{fig:SSAFeynman}. We introduce a Pauli coupling besides the conventional Dirac coupling at the vertex between the struck quark and the virtual photon~\cite{Kochelev:2003cp}:
\be \label{eq:vertexpqq}
V^\gamma_\mu=\mcf_D^q \g^{\mu} - \frac{\mcf_P^q}{2m_q} \sigma^{\mu\nu} q_\nu
\ee
with $m_q$ being the constitute quark mass. The $\mcf_D^q$ and $\mcf_p^q$ are Dirac and Pauli form factors as a function of photon virtuality $Q^2 = - q^2$, respectively. The perturbative and non-perturbative QCD contributions to the Dirac and Pauli form factor have been calculated with several methods (see~\cite{Kochelev:2003cp,Dorokhov:2002kr} and references therein). It should be noted that the non-perturbative interaction enters into the model only through above quark-photon vertex. It does not affect the kinematical relationships of involved momenta, but it would influence the virtualities of internal lines of hard scattering. As a matter of fact, it can be easily checked that the Pauli coupling $\mcf_P^q Q/m_q$ present in Eq.~(\ref{eq:vertexpqq}) is very moderately dependent on $Q^2$ and roughly follow the order of the usual Diral vector coupling $\mcf_D^q$ in the instanton liquid model because of the small instanton size (see Eq.~(\ref{eq:pauliqP}) hereafter in Sec.~\ref{sec:numer}). So it is naively expected this new vertex does not break the TMD factorization between the hard and soft part of SIDIS, and the standard treatment of the convolution between TMD PDFs and FFs can be still used here in the factorization framework~\cite{Ji:2004wu,Bacchetta:2006tn}. Moreover, in our previous calculation of the SSA in SIDIS induced by the Pauli couplings~\cite{Cao:2017bdi}, terms which obviously violate order counting in TMD factorization are not present. There we also found that the Pauli coupling in quark-gluon vertex from instanton liquid model does provide appropriate regularization of the loop integrals from final state interaction. In Ref. [26], the Sivers function resulted from this Pauli coupling is successfully calculated by the correlator in terms of gauge links, adopting the same instanton model with ours and the quark wave functions from MIT bag model. A rigorous proof of the TMD factorization theorem with regard to the Pauli couplings would follow the framework recently proposed within the scalar diquark model in our interested kinematical range and is waiting for future exploration~\cite{Moffat:2017sha}.

The nucleon-quark-diquark vertices are chosen to be
\be
V_s = i g^s \textbf{1} \quad, \qquad
V_{\mu}^a = i \frac{g^a}{\sqrt{2}} \gamma_5 \gamma^\mu \quad,
\ee
where the $g^s$ and $g^a$ are couplings in proton-quark-scalar diquark vertex and proton-quark-axial-vector diquark vertex, respectively. Then the amplitudes of Born diagram presented in Fig.~\ref{fig:SSAFeynman} are straightforwardly written as
\bea
i \mca^\lambda_{s,s'} &=& \frac{- i}{r^2+m_q^2} \bar{u}(q+r,s')\epsilon^\lambda_\mu(q) \lf(\mcf_D^q \g^{\mu} - \frac{\mcf_P^q}{2m_q} \sigma^{\mu\nu} q_\nu \rg) (\rslash+m_q) u(P,s) \quad\quad\\
i \mca^{\lambda,\lambda_a}_{s,s'} &=& \frac{- i}{r^2+m_q^2} \bar{u}(q+r,s')\epsilon^\lambda_\mu(q) \lf(\mcf_D^q \g^{\mu} - \frac{\mcf_P^q}{2m_q} \sigma^{\mu\nu} q_\nu\rg) (\rslash+m_q) \nonumber \\ && \times \frac{\g_5}{\sqrt{2}} \kern 0.2 em \varepsilon\kern -0.50em/_D(P-r,\lambda_a) u(P,s) \quad \quad
\eea
for the scalar and axial-vector diquark models, respectively. Here the $\lambda$ and $\lambda_a$ are helicities of the photon and axial-vector diquark, respectively. The $s$ and $s'$ label the helicities of initial and final quarks, respectively. The helicity amplitudes are obeying the relations
\be
\mca_{s,s'}^{\lambda(,\lambda_a)} = -(-1)^{s-s'} \left(\mca_{-s,-s'}^{-\lambda(,-\lambda_a)}\right)^*
\ee

Following the approach of Hoyer and Jarvinen~\cite{Hoyer:2005ev}, we work in a coordinate system where the target proton is at rest (laboratory frame) and the virtual photon momentum is along the +z axis, e.g. the momenta of photon, proton and struck quark read~\cite{Hoyer:2005ev}
\bea \label{eq:cmsystem}
 q &=& (q^+,q^-,\vec{0}_\perp) \simeq (2\nu,-x M,\vec{0}_\perp)
 \nonumber \\
 P &=& (P^+,P^-,\vec{0}_\perp) \simeq (M,M,\vec{0}_\perp)
 \nonumber \\
 r &=& (r^+,r^-,\vec{r}_\perp) \simeq (xM, xM, r_\perp \cos\phi, r_\perp \sin\phi)
\eea
with $\nu = Q^2/2xM$ being the photon energy, $M$ the proton mass and $x$ the Bjorken variable, respectively. Here we describe a generic four-vector $k$ as $k = (k^+,k^-,\vec{k}_\perp)$ and its square $k^2 = (k^+k^-+k^-k^+)/2 + \vec{k}_\perp^2$. We use $k_\perp$ as the magnitude of $\vec{k} _\perp$. The polarization vectors of photon and axial-vector diquark are defined as
\bea
\epsilon^\lambda(q) &=& \frac{1}{\sqrt{2}} \lf(0,\,0, -\lambda, - i\rg)
\\
\epsilon_D(P-r,\lambda_a) &=& \frac{1}{\sqrt{2}} \lf(\frac{2\,(\lambda_a\,r_x + i\,r_y)}{(1-x)\,M}, 0, -\lambda_a, - i\rg)
\eea
where the transversality condition, e.g. $(P-r) \cdot \epsilon_D(P-r,\lambda_a) = 0$, is respected. For simplicity, we define,
\be
\mca_{s,s'}^{\lambda(,\lambda_a)} = g_{s,a} \sqrt{2M q^+} \, \frac{1-x}{\vec{r}_\perp^2+B_R^2(m_q^2)} \,\mci_{s,s'}^{\lambda(,\lambda_a)}
\ee
with $B_R^2(m_q^2) = (1-x)m_q^2 +x m_D^2 - x(1-x) M^2$ and $m_D$ being the diquark mass. After some algebraic caculation, in the limit of $q^+ = Q^2/xM \to\infty$ at fixed $k,r$, the helicity amplitudes for the scalar diquark model are found to be
\bea
\mci^{\lambda}_{+,+} &\simeq& \lf(\mcf_D^q - \frac{\mcf_P^q}{2m_q} D_Q\rg) \, r_\perp e^{+i\psi}\, \delta_{\lambda,+1} + \frac{\mcf_P^q}{2m_q}\, D_R \, r_\perp e^{-i\psi} \,\delta_{\lambda,-1} \label{eq:Apps}\\
\mci^{\lambda}_{+,-} &\simeq& -\frac{\mcf_P^q}{2m_q} \vec{r}_\perp^2 e^{2i\psi} \, \delta_{\lambda,+1} + \lf( \mcf_D^q + \frac{\mcf_P^q}{2m_q} D_R \rg) D_R \, \delta_{\lambda,-1} \label{eq:Apms}
\eea
with $D_Q = x M-m_q$ and $D_R = x M+m_q$. For the axial-vector diquark model, the full set of amplitudes is
\bea
\mci^{+,+}_{+,+} &\simeq& - \lf( \mcf_D^q + \frac{\mcf_P^q}{2m_q} D_R \rg) D_R - \frac{\mcf_P^q}{2m_q} \frac{x}{1-x} \,\vec{r}_\perp^2 \label{eq:Appppsv} \\
\mci^{+,-}_{+,+} &\simeq& \frac{\mcf_P^q}{2m_q} \frac{x}{1-x} \,\vec{r}_\perp^2 e^{+2i\psi} \label{eq:Apmppv} \\
\mci^{-,+}_{+,+} &\simeq& \frac{\mcf_P^q}{2m_q} \frac{1}{1-x} \,\vec{r}_\perp^2 e^{-2i\psi} \\
\mci^{-,-}_{+,+} &\simeq& - \frac{\mcf_P^q}{2m_q} \frac{x}{1-x} \, \vec{r}_\perp^2 \\
\mci^{+,+}_{+,-} &\simeq& \frac{\mcf_P^q}{2m_q} \, D_R \, r_\perp e^{+i\psi} \\
\mci^{+,+}_{-,+} &\simeq& - \lf(\mcf_D^q + \mcf_P^q \rg) \frac{x}{1-x} r_\perp e^{-i\psi} \\
\mci^{+,-}_{-,+} &\simeq& \lf[ \mcf_D^q  + \frac{\mcf_P^q}{2m_q} \lf( x\,D_R - D_Q \rg) \rg] \frac{r_\perp e^{+i\psi}}{1-x} \\
\mci^{+,-}_{+,-} &\simeq& 0
\eea

The differential cross section of unpolarized and longitudinal polarized DIS is given in terms of the amplitude at tree order in Fig~\ref{fig:SSAFeynman}(a):
\bea \label{eq:disdif}
\frac{\textrm{d} \sigma}{\textrm{d} x\,\textrm{d} y\,\textrm{d}^2\vec{r}_{\bot}}  &=& \frac{1}{Q^4} \sum^{(\lambda_a,\lambda_a')}_{\lambda,\lambda',s,s'} \lf\{L^{\lambda,\lambda'} \mca^{\lambda(,\lambda_a)}_{s,s'} \lf( \mca^{\lambda'(,\lambda_a')}_{s,s'} \rg)^* - S_{\|} \sqrt{1-\varepsilon^2}\,\lambda\, L^{\lambda,\lambda'} \mca^{\lambda(,\lambda_a)}_{s,s'} \lf( \mca^{\lambda'(,\lambda_a')}_{s,s'} \rg)^* \rg\} \nonumber
\\ \noindent &=& \frac{16 e^2 g_{s,a}^2}{x\,y^2} \frac{y^2}{2\,(1-\varepsilon)} \lf( 1 + \frac{\gamma^2}{2\,x}\rg) \lf(\frac{1-x}{\vec{r}_{\bot}^2 +B_R^2(m_q^2)}\rg)^2 \left[ \mcn_{+} - \varepsilon\,\mcn_{-} + S_{\|} \sqrt{1-\varepsilon^2} \mcn_{\|}\right] \qquad
\eea
with the Bjorken variables $y$ being the fraction of the beam energy carried by the virtual photon. The depolarization factor $\varepsilon$ is the ratio of longitudinal and transverse photon flux,
\be
\varepsilon = \frac{1-y -\frac{1}{4}\slim \gamma^2 y^2}{1-y+\frac{1}{2}\slim y^2 +\frac{1}{4}\slim \gamma^2 y^2}
\ee
with $\gamma = 2x M/ Q$. Above we have used the leptonic tensor in the helicity basis
\bea \label{eq:Lexpr}
L^{\lambda,\lambda'} &=& \frac{4e^2 Q^2}{y^2} \frac{y^2}{2\,(1-\varepsilon)} \lf( 1 + \frac{\gamma^2}{2\,x}\rg) \left\{\delta_{\lambda,\lambda'} - \varepsilon e^{-2i\lambda\tau}\delta_{\lambda,-\lambda'} \right\}
\eea
Here $\tau$ is the azimuthal angle of the lepton $\textbf{\emph{l}}_{1\perp}=\textbf{\emph{l}}_{2\perp}=(l_\perp \cos\tau, l_\perp \sin\tau)$. After some formalism manipulation, we have for the scalar diquark model
\bea \label{eq:calcNscalarp}
\mcn_{+}^s &=& \sum_{\lambda} \lf\{ |\mca^{\lambda}_{+,+}|^2 + |\mca^{\lambda}_{+,-}|^2 \rg\} \nonumber \\ &\simeq& \lf(\mcf_D^q - \frac{\mcf_P^q}{2m_q} D_Q \rg)^2 \,\vec{r}_{\bot}^2 + \lf(\mcf_D^q + \frac{\mcf_P^q}{2m_q} D_R \rg)^2 \,D_R^2 + \lf(\frac{\mcf_P^q}{2m_q}\rg)^2 \,\vec{r}_{\bot}^2 \,(\vec{r}_{\bot}^2+D_R^2) \quad
\\ \nonumber &\xrightarrow{\mcf_P^q \sim 0}& (\mcf_D^q)^2 ( \vec{r}_{\bot}^2 + D_R^2 )
\\ \mcn_{-}^s &=& \mathrm{Re} \sum_{\lambda} \lf\{ \mca^\lambda_{+,+} \lf(\mca^{-\lambda}_{+,+}\rg)^* + \mca^\lambda_{+,-} \lf(\mca^{-\lambda}_{+,-}\rg)^* \rg\}e^{-2i\lambda\tau} \nonumber \\ &\simeq& -\lf(\frac{\mcf_P^q}{2m_q}\rg)^2\,2 x M D_R \vec{r}_{\bot}^2 \cos2(\psi-\tau) \label{eq:calcNcahns}
\\ \mcn_{\|}^s &=& - \sum_{\lambda} \lambda\, \lf\{ |\mca^{\lambda}_{+,+}|^2 + |\mca^{\lambda}_{+,-}|^2 \rg\} \nonumber \\ &\simeq& - \lf(\mcf_D^q - \frac{\mcf_P^q}{2m_q} D_Q \rg)^2 \,\vec{r}_{\bot}^2 + \lf(\mcf_D^q + \frac{\mcf_P^q}{2m_q} D_R \rg)^2 \,D_R^2 - \lf(\frac{\mcf_P^q}{2m_q}\rg)^2 \,\vec{r}_{\bot}^2 \,(\vec{r}_{\bot}^2-D_R^2) \qquad \label{eq:calcNscalarm}
\\ \nonumber &\xrightarrow{\mcf_P^q \sim 0}& (\mcf_D^q)^2 (D_R^2 - \vec{r}_{\bot}^2)
\eea
For axial-vector diquark model, we have
\bea \nonumber
\mcn_+^a &=& \sum_{\lambda,\lambda_a} \lf\{ |\mca^{\lambda,\lambda_a}_{+,+}|^2 + |\mca^{\lambda,\lambda_a}_{+,-}|^2 \rg\} \\ &&
\begin{aligned}
\simeq \lf[ \lf( \mcf_D^q + \frac{\mcf_P^q}{2m_q} D_R \rg) D_R + \frac{\mcf_P^q}{2m_q} \frac{x}{1-x} \,\vec{r}_\perp^2 \rg]^2 + \qquad\qquad\qquad\qquad \\ \frac{\vec{r}_\perp^2}{(1-x)^2} \lf[ \lf(\mcf_D^q + \mcf_P^q \rg)^2\,x^2 + \lf( \mcf_D^q  + \frac{\mcf_P^q}{2m_q} \lf( x\,D_R - D_Q \rg) \rg)^2 \rg. \\ \lf. + \lf(\frac{\mcf_P^q}{2m_q} \rg)^2 \,\lf((1+2\,x^2)\,\vec{r}_\perp^2 + (1-x)^2\,D_R^2\rg) \rg] \qquad \label{eq:calcNaxialp}
\end{aligned} \\ \nonumber
&\xrightarrow{\mcf_P^q \sim 0}& (\mcf_D^q)^2 \lf(D_R^2 + \frac{1+x^2}{(1-x)^2} \vec{r}_\perp^2 \rg) \\
\mcn_-^a &=& \mathrm{Re} \sum_{\lambda,\lambda_a} \lf\{ \mca^{\lambda,\lambda_a}_{+,+} \lf(\mca^{-\lambda,-\lambda_a}_{+,+}\rg)^* + \mca^{\lambda,\lambda_a}_{+,-} \lf(\mca^{-\lambda,-\lambda_a}_{+,-}\rg)^* \rg\}e^{-2i\lambda\tau} \nonumber
\\ &\simeq&  -\lf(\frac{\mcf_P^q}{2m_q}\rg)^2\,\frac{\vec{r}_\perp^2}{1-x} \, \lf[ (1-x)D_R^2 +D_R D_Q + \frac{\vec{r}_\perp^2}{1-x} x(1+x) \rg] \cos2(\psi-\tau) \label{eq:calcNcahnv} \nonumber
\eea
\bea
\mcn_{\|}^a &=& - \sum_{\lambda,\lambda_a} \lambda \lf\{ |\mca^{\lambda,\lambda_a}_{+,+}|^2 + |\mca^{\lambda,\lambda_a}_{+,-}|^2 \rg\} \\ && \begin{aligned}
\simeq - \lf[ \lf( \mcf_D^q + \frac{\mcf_P^q}{2m_q} D_R \rg) D_R + \frac{\mcf_P^q}{2m_q} \frac{x}{1-x} \,\vec{r}_\perp^2 \rg]^2 + \qquad\qquad\qquad\qquad \\ \frac{\vec{r}_\perp^2}{(1-x)^2} \lf[ \lf(\mcf_D^q + \mcf_P^q \rg)^2\,x^2 + \lf( \mcf_D^q  + \frac{\mcf_P^q}{2m_q} \lf( x\,D_R - D_Q \rg) \rg)^2 \rg. \\ \lf. + \lf(\frac{\mcf_P^q}{2m_q} \rg)^2 \,\lf(\vec{r}_\perp^2 - (1-x)^2\,D_R^2\rg) \rg] \qquad \label{eq:calcNaxialm}
\end{aligned} \\ \nonumber
&\xrightarrow{\mcf_P^q \sim 0}& (\mcf_D^q)^2 \lf( \frac{1+x^2}{(1-x)^2} \vec{r}_\perp^2 - D_R^2 \rg)
\eea
As expected, the distributions $\mcn_+^{s,a}$ and $\mcn_{\|}^{s,a}$ return back to the results of conventional diquark models~\cite{Bacchetta:2008af} when $\mcf_P^q$ is approaching zero. Besides, an asymmetric term $\mcn_-^{s,a}$ with definite angular distribution are generated by the photon-quark Pauli couplings. If we use the Trento convention \cite{Bacchetta:2004jz}, which uses the angle $\phi_h$ between the hadron $h$ and lepton planes
\be \label{eq:Trentopsi}
 \phi_h = \psi-\tau
\ee
Then the angular dependence of $\mcn_-^{s,a}$ are $\cos2(\psi-\tau) = \cos 2\phi_h$, which is identical to those of the Cahn and Boer-Mulders effect. Its physical origin is, however, quite different from the original $\cos 2\phi_h$ azimuthal asymmetries in SIDIS. As can be seen in Eq.~(\ref{eq:calcNcahns}) and Eq.~(\ref{eq:calcNcahnv}), the magnitudes of $\mcn_-^{s,a}$ are proportional to the squared Pauli form factor $(\mcf_P^q)^2$, which is from non-perturbative contribution and could be explicitly calculated by various phenomenological models, e.g. the instanton liquid model for nontrivial topological structure of QCD vacuum~\cite{Kochelev:2003cp}. It should be noted that the Pauli couplings $\mcf_P^q$ introduced in Eq.~(\ref{eq:pauliqP}) are $Q^2$ dependent, which will discussed in detail in Sec.~\ref{sec:numer}.

In our above calculation, we neglect the form factors which are introduced to smoothly suppress the contribution of high transverse momentum. Here we introduce conventional dipolar form factors in the photon-quark couplings:
\be \label{eq:formfactor}
\mcf_{D,P}^q \Longrightarrow \mcf_{D,P}^q \frac{r^2-m_q^2}{(r^2-\Lambda_{s,a}^2)^2} = \mcf_{D,P}^q \, \frac{\vec{r}_{\bot}^2 + B_R^2(m_q^2)}{(\vec{r}_{\bot}^2 +B_R^2(\Lambda_{s,a}))^2}\,(x-1)
\ee

We can read the unpolarized and helicity parton distribution functions (PDFs) from Eq.~(\ref{eq:calcNscalarp}), Eq.~(\ref{eq:calcNscalarm}), Eq.~(\ref{eq:calcNaxialp}), and Eq.~(\ref{eq:calcNaxialm}):
\bea \label{eq:f1kts}
f_1^{s}(x,\vec{r}_{\bot}) &=& \frac{g_s^{2}(1-x)^3}{(2\pi)^3} \frac{\mcn_+^s}{2\,(\vec{r}_{\bot}^2 +B_R^2(\Lambda_s^2))^4}
\\ \label{eq:f1kta}
f_1^{a}(x,\vec{r}_{\bot}) &=& \frac{g_a^{2}(1-x)^3}{(2\pi)^3} \frac{\mcn_+^a}{2\,(\vec{r}_{\bot}^2 +B_R^2(\Lambda_a^2))^4}
\\ \label{eq:g1kts}
g_{1L}^{s}(x,\vec{r}_{\bot}) &=& \frac{g_s^{2}(1-x)^3}{(2\pi)^3} \frac{\mcn_{\|}^s}{2\,(\vec{r}_{\bot}^2 +B_R^2(\Lambda_s^2))^4}
\\ \label{eq:g1kta}
g_{1L}^{a}(x,\vec{r}_{\bot}) &=& \frac{g_a^{2}(1-x)^3}{(2\pi)^3} \frac{\mcn_{\|}^a}{2\,(\vec{r}_{\bot}^2 +B_R^2(\Lambda_a^2))^4}
\eea
which return back to the results of conventional diquark model~\cite{Hoyer:2005ev} when $\mcf_P^q \rightarrow 0$. We can also define a new distribution function alike to Cahn effect by Eq.~(\ref{eq:calcNcahns}) and Eq.~(\ref{eq:calcNcahnv}) as
\bea \label{eq:Cahnlikekts}
{\mathcal C}^s(x,\vec{r}_{\bot}) &=& - \frac{g_s^{2}(1-x)^3}{(2\pi)^3} \frac{\mcn_-^s}{2\,(\vec{r}_{\bot}^2 +B_R^2(\Lambda_s^2))^4}
\\ \label{eq:Cahnlikekta}
{\mathcal C}^a(x,\vec{r}_{\bot}) &=& - \frac{g_a^{2}(1-x)^3}{(2\pi)^3} \frac{\mcn_-^a}{2\,(\vec{r}_{\bot}^2 +B_R^2(\Lambda_a^2))^4}
\eea
In above equations, $g_{s,a}$ are the normalization factors determined by
\be
\int_0^1 \textrm{d}x \int_0^\infty \textrm{d}^2 \vec{r}_{\bot}\, f_1^{s,a} (x,\vec{r}_{\bot}) = \int_0^1 \textrm{d} x\, f_1^{s,a} (x) = 1
\ee
The $\vec{r}_{\bot}$-integrated results $f_1^{s,a} (x)$, $g_{1L}^{s,a}(x)$ and ${\mathcal C}^{s,a}(x)$ are given in the Appendix~\ref{apdx:rt}.

\begin{table*}[t]
\begin{center}
\begin{tabular}{|c|c|c|}
\hline & {Fit-I} & {Fit-II} \\
\hline
$m_q$ (GeV)             & 0.3$^\dag$          & 0.3$^\dag$         \\
$m_D^{s}   $ (GeV)      & 0.769 $\pm$ 0.038   & 0.829 $\pm$ 0.025  \\
$m_D^{a(u)}   $ (GeV)   & 1.5$^\dag$          & 1.5$^\dag$         \\
$m_D^{a(d)}   $ (GeV)   & 0.857 $\pm$ 0.065   & 0.921 $\pm$ 0.007  \\
$\Lambda_{s}$ (GeV)     & 0.683 $\pm$ 0.028   & 0.545 $\pm$ 0.017  \\
$\Lambda_{a(u)}$ (GeV)  & 0.767 $\pm$ 0.030   & 0.509 $\pm$ 0.025  \\
$\Lambda_{a(d)}$ (GeV)  & 0.434 $\pm$ 0.027   & 0.385 $\pm$ 0.006  \\
$c_{s}^2$               & 0.445 $\pm$ 0.048   & 0.657 $\pm$ 0.061  \\
$c_{a(u)}^2$            & 1.245 $\pm$ 0.060   & 1.112 $\pm$ 0.089  \\
$c_{a(d)}^2$            & 0.639 $\pm$ 0.011   & 0.662 $\pm$ 0.012  \\
$\mcf_P^q(Q^2_0)$        &        0.0$^\dag$  & 0.921 $\pm$ 0.117  \\
$\chi^2/d.o.f$          &       8.51          &       4.12         \\
\hline
\end{tabular}
\end{center}
\caption{Extracted parameters in two fit schemes. The $^\dag$ labels the parameter with a fixed value in the fit.
\label{Tab:fitpara}}
\end{table*}

After determining the parameters of the model by fitting known unpolarized and polarized distribution functions $f_1^{u,d} (x)$ and $g_1^{u,d} (x)$, we can predict the numerical results of our model for the $\cos 2\phi_h$ asymmetry.  To perform the fit, we need to relate the functions $f_1^{s,a} (x)$ and $g_1^{s,a} (x)$ in terms of diquark types in the model, to the functions $f_1^{u,d} (x)$ and $g_1^{u,d} (x)$ with regard to quark flavors in the global fits. We write them generally as~\cite{Bacchetta:2003rz,Bacchetta:2008af}
\bea
f_1^{u} &=& c_{s}^2 f_1^{s} + c_{a(u)}^2 f_1^{a(u)} \\
f_1^{d} &=& c_{a(d)}^{2} f_1^{a(d)}
\eea
which should be generalized to the case of $g_1^{u,d}$ and ${\mathcal C}^{u,d}$. Here we discriminate the two isospin states of the vector diquark, namely $a(u)$ for the $ud (I_3 = 0)$ diquark and $a(d)$ for the $uu (I_3 = 1)$ diquark. We have $c_{s}^2 = 3/2$, $c_{a(u)}^2 = 1/2$ and $c_{a(d)}^{2} = 1$ under the SU(4) symmetry. However, since SU(4) symmetry is no longer strictly preserved in the spectator model of nucleon, we treat these coefficients instead as free parameters. We have checked that the results are not sensitive to the values of the masses of constituent quark $m_q$ and $ud$ diquark $m_D^{a(u)}$, so we fix them to be $m_q = $ 0.30 GeV and $m_D^{a(u)} = $  1.5 GeV. We have in total nine free parameters for the model.

In order to reproduce the parametrizations of PDFs extracted from experimental data, we have to choose a proper scale $Q^2_0$ at which diquark models are probably appropriate. Following the arguments in literatures~\cite{Bacchetta:2003rz,Bacchetta:2008af}, a very low scale $Q^2_0$ seems to be favored, so lowest possible value of scale is used in the literature, e.g. $0.3$ GeV$^2$~\cite{Bacchetta:2008af} and $0.078$ GeV$^2$~\cite{Bacchetta:2003rz}. Only a few PDFs of global fit are at hand in such low $Q^2$ range and most of them are usually applicable above 1.0 GeV$^2$,  e.g. CT14~\cite{Dulat:2015mca}, MMHT2014~\cite{Harland-Lang:2014zoa}, NNPDF~\cite{Ball:2014uwa}, and ABM11~\cite{Alekhin:2012ig}. We fit our model parameters to leading order GRV1998~\cite{Gluck:1998xa} and GRSV2000~\cite{Gluck:2000dy} parametrization at $Q^2_0$ = 0.26 GeV$^2$ for $f_1^{u,d}$ and $g_1^{u,d}$, respectively. We assign a constant relative error of 5.0\% to $f_1^{u,d}$ based on comparisons with HERAPDF~\cite{Abramowicz:2015mha} and IMParton~\cite{Chen:2013nga,Chen:2013oga} under the same $Q^2$. We allocate a relative error of 20.0\% to $g_1^{u,d}$ in view of confronting with DNS2005~\cite{deFlorian:2005mw} and DSSV2010~\cite{deFlorian:2009vb} (see Ref~\cite{Hirai:2003pm,Shahri:2016uzl} for a detailed comparison of various $g_1^{u,d}$).

\begin{figure}
\begin{center}
{\includegraphics*[width=12.cm]{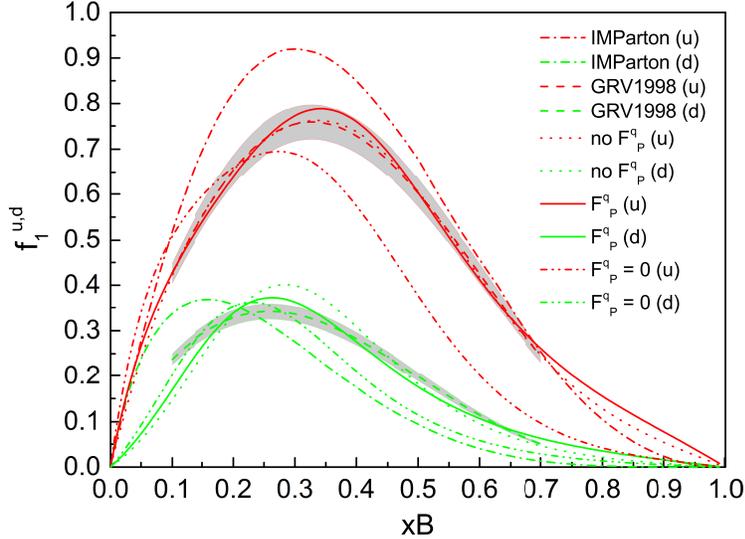}}
\caption{(color online) The unpolarized PDF $f_1^{u,d}$ in our model . The red and blue lines are for $u-$ and $d-$quark, respectively.
  The dashed and dash-dotted lines are from GRV1998~\cite{Gluck:1998xa} and IMParton~\cite{Chen:2013nga,Chen:2013oga}, respectively. The shadow bands correspond to the relative error of 5.0\% to GRV1998. The dotted and solid curve are the results of Fit-I and Fit-II, respectively. The dash-dot-dotted curves represent result of Fit-II with setting the Pauli coupling in photon-guark vertex to be zero (${\mcf_P^q} = 0$).
\label{fig:comPDF}}
\end{center}
\end{figure}

\begin{figure}
\begin{center}
{\includegraphics*[width=12.cm]{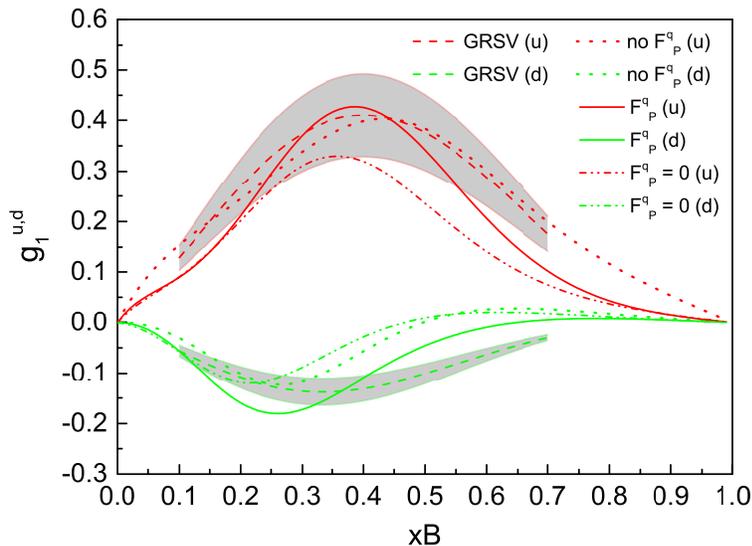}}
\caption{(color online) The polarized PDF $g_1^{u,d}$ in our model. The red and blue lines are for $u$ and $d$ quark, respectively.
  The dashed lines are from GRSV2000~\cite{Gluck:2000dy} and the shadow bands correspond to their relative error of 20.0\%. Other curves are the same with the labels in Fig.~\ref{fig:comPDF}.
\label{fig:compPDF}}
\end{center}
\end{figure}

The fit results are shown in Fig.~\ref{fig:comPDF} and Fig.~\ref{fig:compPDF} for $f_1^{u,d}$ and $g_1^{u,d}$, respectively. The extracted parameters are shown in Tab.~\ref{Tab:fitpara}. In Fit-I, we use the conventional model, in which there is no Pauli coupling in photon-guark vertex, e.g setting ${\mcf_P^q} = 0$ in the fit. We achieve a good agreement except for the $g_1^{d}$. The fitted $g_1^{d}$ goes to positive values in the range of around $x = 0.6$, while it is negative in the full $x$ range in GRSV2000. This conclusion has been found before in Ref.~\cite{Bacchetta:2008af}. In Fit-II, we use the full model with ${\mcf_P^q}$ at $Q^2_0$ = 0.26 GeV$^2$ as free parameter. In spite of the high $\chi^2$ because we choose small relative errors for $f_1^{u,d}$, the agreement is acceptable for all the PDFs and the overall $\chi^2$ is promoted significantly. The fit of $f_1^{u}$ is as good as that in Fit-I, and the $f_1^{d}$ is slightly improved. Though the description of $g_1^{u}$ becomes a little worse in Fit-II, the $g_1^{d}$ is much better from $x = 0.3$ to $0.7$. The contribution of Pauli coupling is significant above $x = 0.3$, as can be seen in Fig.~\ref{fig:comPDF} and Fig.~\ref{fig:compPDF}, when we turn off it in the fit results of Fit-II.

\begin{figure}
\begin{center}
{\includegraphics*[width=10.cm]{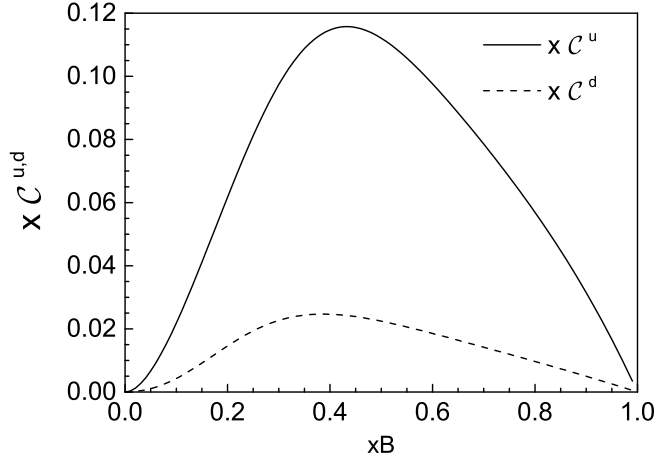}}
\caption{The distributions $x\,{\mathcal C}^{u,d}(x)$ at $Q^2_0 = 0.26$ GeV$^2$ in our model. The solid and dashed lines are for $u$ and $d$ quark, respectively.
\label{fig:cahnlike}}
\end{center}
\end{figure}

In Fig.~\ref{fig:cahnlike}, we give the predicted distributions $x\,{\mathcal C}^{u,d}(x)$ at $Q^2_0 = 0.26$ GeV$^2$ in our model. The magnitude of $u-$quark distribution is much bigger than that of $d-$quark, though the their shape is close to each other. The maximum for both distributions is around at $x \simeq 0.4$.

\section{Comparism to data of $\cos 2\phi_h$ asymmetry} \label{sec:numer}

The $\cos 2\phi_h$ asymmetries in SIDIS off hydrogen, deuterium and $^3$He targets have already been extensively explored by EMC~\cite{Arneodo:1986cf}, ZEUS~\cite{Breitweg:2000qh,Chekanov:2006gt}, HERMES~\cite{Giordano:2010gq,Airapetian:2012yg}, COMPASS~\cite{Collaboration:2010fi,Kafer:2008ud,Bressan:2009eu,Adolph:2014pwc} collaborations and at JLab~\cite{Mkrtchyan:2007sr,Osipenko:2008aa,Yan:2016ods}. Since both HERMES~\cite{Airapetian:2012yg} and COMPASS~\cite{Adolph:2014pwc} collaborations have measured the SIDIS off unpolarized proton for charged unidentified hadrons ($h$) among final particles but under slightly different $Q^2$, here we try to compare our new asymmetry in Eq.~(\ref{eq:Cahnlikekts}) and Eq.~(\ref{eq:Cahnlikekta}) with the parameters in Tab.~\ref{Tab:fitpara} to these data, together with Cahn effect. We do not aim to a full analysis of the azimuthal asymmetry based on all released data, because it is beyond the scope of present paper. Attempts to this direction can be found in a series of publications~\cite{Barone:2005kt,Barone:2008tn,Barone:2009hw,Barone:2015ksa,Zhang:2008ez,Zhang:2008nu,Lu:2009ip}. For the same reason, we also do not investigate Boer-Mulders distribution~\cite{Boer:1997nt} here, for which Collins FFs of the production hadron should be introduced~\cite{Collins:1992kk,Collins:2002kn}.

In parton model, the SIDIS differential cross sections are expressed as the convolution of TMD PDFs $f_1(x,\vec{r}_{\bot})$ and FFs $D_1^q(z_h,\vec{p}_{\bot})$~\cite{Barone:2005kt,Barone:2008tn,Barone:2009hw,Barone:2015ksa}:
\bea
\frac{d \sigma}{d x\,d y\,d z_h\,d P_{hT}^2\,d \phi_h}  &=& \frac{2\,\pi\,\alpha^2}{Q^4\,x\,y} \sum_{q} e_q^2\,x \frac{y^2}{2\,(1-\varepsilon)} \lf( 1 + \frac{\gamma^2}{2\,x}\rg) \nonumber \\ &\times& \int \textrm{d}^2\vec{r}_{\bot} \textrm{d}^2 \vec{p}_{\bot} \delta^2(\vec{P}_{hT}-z\,\vec{r}_{\bot}-\vec{p}_{\bot}) f^q_1(x,\vec{r}_{\bot}) D_1^q(z_h,\vec{p}_{\bot}) \qquad
\eea
where the sum runs over the quark favor $q$. We denote by $\vec{P}_{hT}$ the transverse momentum of the final hadron $h$, and by $\vec{p}_{\bot}$ the transverse momentum of $h$ with respect to the direction of the fragmenting quark. The $D_1^q(z_h,\vec{p}_{\bot})$ is well recognized in the Gaussian form:
\be
D_1^q(z_h,\vec{p}_{\bot}) = D_1^q(z_h) \frac{e^{-\vec{p}_{\bot}^2/\langle\vec{p}_{\bot}^2\rangle}}{\pi \langle\vec{p}_{\bot}^2\rangle}
\ee
We adopt the unpolarized DSS FFs at leading order for $D_1^q(z_h)$~\cite{deFlorian:2007aj} with $\langle\vec{p}_{\bot}^2\rangle =$ 0.2 GeV$^2$~\cite{Anselmino:2005nn,Barone:2015ksa}. The results do not change much if the SKMA FFs~\cite{Soleymaninia:2013cxa} are used instead.

Similarly, the contribution of the distribution ${\mathcal C}^{q}(x,\vec{r}_{\bot})$ to the $\cos 2\phi_h$ asymmetry can be defined as,
\bea
\frac{d \sigma}{\textrm{d} x\,\textrm{d} y\,\textrm{d} z_h\,\textrm{d} P_{hT}^2\,\textrm{d} \phi_h}  &=& \frac{2\,\pi\,\alpha^2}{Q^4\,x\,y} \sum_{q} e_q^2\,x \frac{y^2\,\varepsilon}{2\,(1-\varepsilon)} \lf( 1 + \frac{\gamma^2}{2\,x}\rg) \nonumber \\ &\times& \int \textrm{d}^2\vec{r}_{\bot} \textrm{d}^2 \vec{p}_{\bot} \delta^2(\vec{P}_{hT}-z\,\vec{r}_{\bot}-\vec{p}_{\bot}) {\mathcal C}^{q}(x,\vec{r}_{\bot}) D_1^q(z_h,\vec{p}_{\bot}) \cos 2\phi_h \qquad
\eea
The Cahn effect to $\cos 2\phi_h$ asymmetry can be obtained by substituting the ${\mathcal C}^{q}(x,\vec{r}_{\bot})$ with the following function ${\mathcal C}_{cahn}^{q}(x,\vec{r}_{\bot})$ in above definition~\cite{Cahn:1978se,Cahn:1989yf}:
\bea
{\mathcal C}_{cahn}^{q}(x,\vec{r}_{\bot}) = 2\,\frac{2\,(\vec{r}_{\bot} \cdott \vec{h})^2 - \vec{r}_{\bot}^2}{Q^2} \,f^q_1(x,\vec{r}_{\bot})
\eea
with $\vec{h} = \vec{P}_{hT}/ |\vec{P}_{hT}|$. As the only known one of the twist-4 effects, it is at the order of $\vec{r}_{\bot}^2/Q^2$ and hence it is controversial when we use the identical kinematical relations and factorization as the leading twist (see discussions in Refs~\cite{Barone:2009hw,Barone:2015ksa}). We have to leave these debate behind at present before these problems are completely resolved from theoretical side.

\begin{figure}
\begin{center}
{\includegraphics*[width=0.8\textwidth]{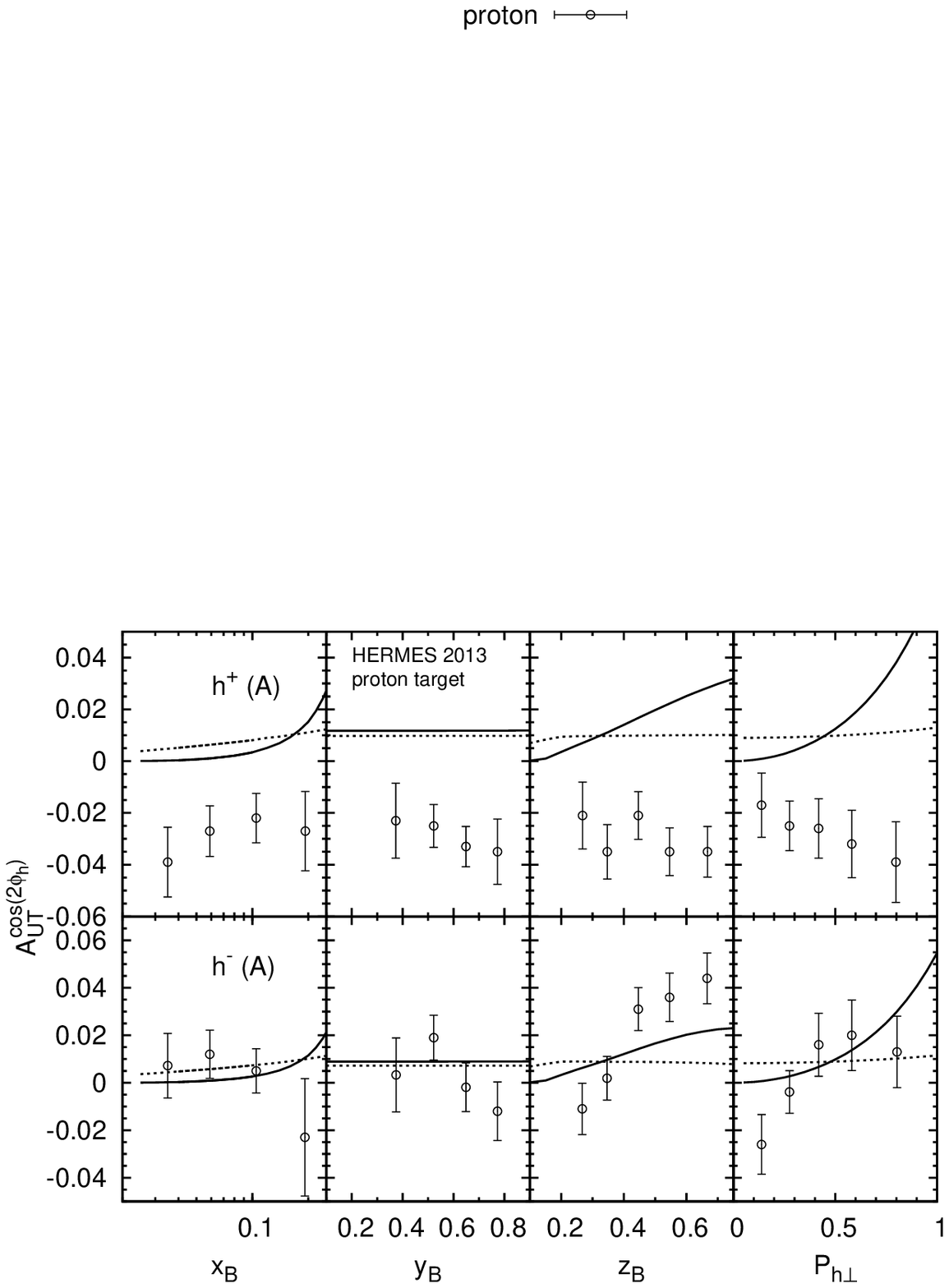}}
\caption{The $\textrm{A}_{\textrm{UT}}^{\cos 2\phi_h}$ in one of kinematic ranges (A) of the HERMES measurement (see TABLE III in Ref.~\cite{Airapetian:2012yg}). The solid and dotted curves are the results of Cahn effect and our new distribution, respectively.
\label{fig:HERMESH}}
\end{center}
\end{figure}

\begin{figure}
\begin{center}
{\includegraphics*[width=0.8\textwidth]{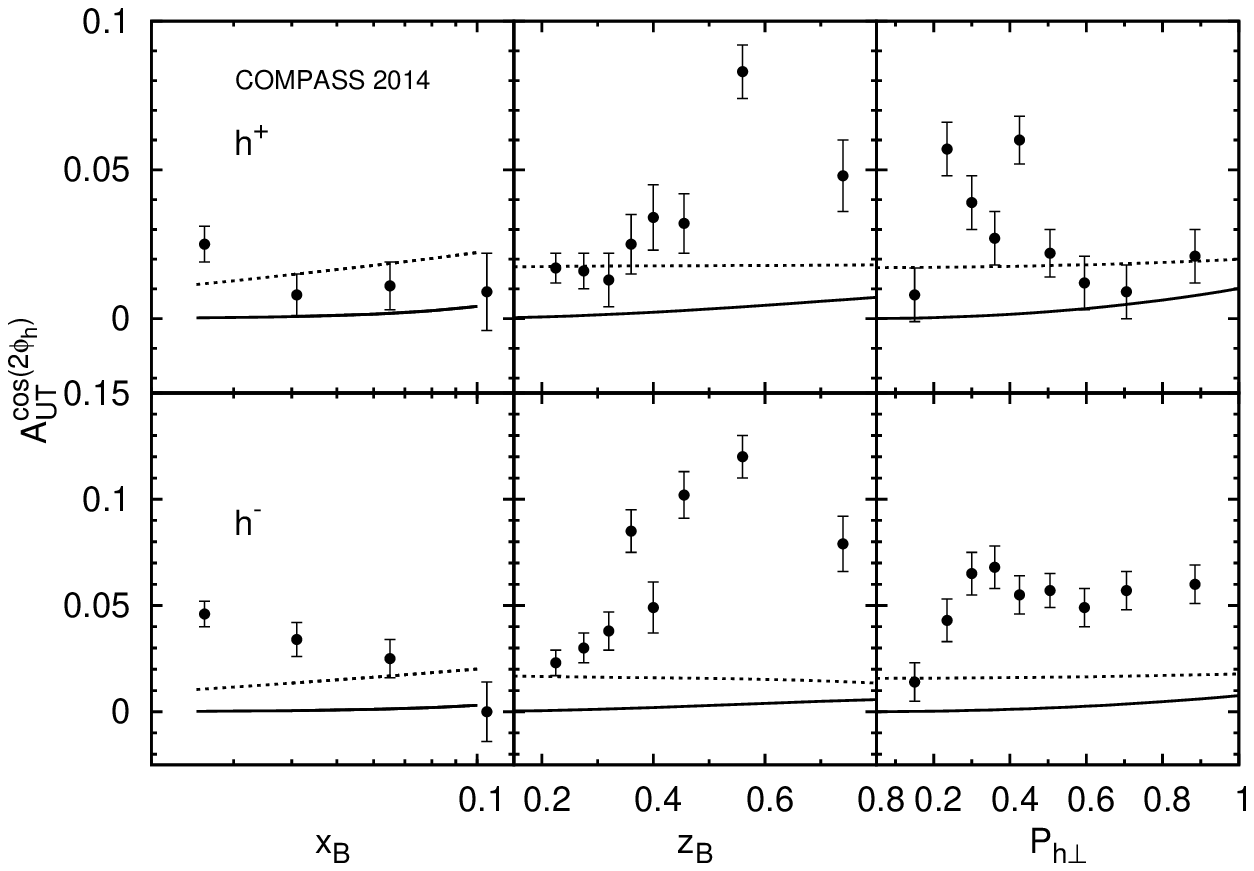}}
\caption{The $\textrm{A}_{\textrm{UT}}^{\cos 2\phi_h}$ in kinematic range of the COMPASS measurement~\cite{Adolph:2014pwc}. The solid and dotted curves are the results of Cahn effect and our new distribution, respectively.
\label{fig:COMPASSH}}
\end{center}
\end{figure}

The extracted value of ${\mcf_P^q}(Q^2_0)$ is close to 1.0 in Sec.~\ref{sec:resul}. This is in fact roughly compatible with the calculation in the instanton model~\cite{Zhang:2017zpi}, where it is found that the $Q^2$ dependence of Pauli coupling can be parameterized very well by
\be \label{eq:pauliqP}
\frac{\mcf_P^q(Q^2)}{\mcf_P^q(Q^2_0)} = \frac{1 + \frac{\rho_c Q^2_0}{4.7 m_q}}{1 + \frac{\rho_c Q^2}{4.7 m_q}}
\ee
with the instanton size $\rho_c = 1/3$ fm. As a result our new distribution is at the order of $1/Q^4$ and would suffer the same kinematical and dynamical problems with the case of Cahn effect as mentioned above. Moreover, the $\mcf_P^q$ decreases very rapidly below $Q^2$ = 1.0 GeV$^2$ and approaches a constant above 2.0 GeV$^2$. As a result, it is expected that the strength of the new asymmetry is enlarged in very low $Q^2$. However, the data in this $Q^2$ range is unavailable. The average $Q^2$ of HERMES~\cite{Airapetian:2012yg} and COMPASS~\cite{Adolph:2014pwc} measurements are about 2.5 and 3.0 GeV$^2$, respectively~\footnote{There is strong correlation between variables $x$ and $Q^2$ (and thus between $x$ and $y$), but we put aside this problem at present and it should pay caution to the sub-figures with $x$ variable in Fig.~\ref{fig:HERMESH} and Fig.~\ref{fig:COMPASSH}}. In principle we need to evolute our distributions from $Q^2_0$ to the experimental one. For the reason that the evolution of these distributions is not firmly established~\cite{Echevarria:2014xaa}, we only simply use the value of $\mcf_P^q(Q^2)$ at experimental $Q^2$, determined by Eq.~(\ref{eq:pauliqP}) during our comparison. This leads to a reduction of about one order of magnitude of ${\mathcal C}^{q}(x,\vec{r}_{\bot})$ compared to that at $Q^2_0$. It should be mentioned here that the data of EMC~\cite{Arneodo:1986cf} and ZEUS~\cite{Breitweg:2000qh,Chekanov:2006gt} are measured in very high $Q^2$, so the contribution of both Cahn effect and our new distribution to $\cos 2\phi_h$ asymmetry are anticipated to be negligible.

The measured asymmetry in experiment is defined as
\be
\textrm{A}_{\textrm{UT}}^{\cos 2\phi_h} = \frac{\int \textrm{d} \sigma \cos 2\phi_h}{\int \textrm{d} \sigma}
\ee
where the integrations are performed over the measured kinematical ranges of $x$, $y$, $z_h$ and $P_{hT}$, which can be found in the papers of HERMES~\cite{Airapetian:2012yg} and COMPASS~\cite{Adolph:2014pwc} collaborations. The detailed formalism of integrations are present widely in the literatures~\cite{Barone:2015ksa,Zhang:2008nu}.

Our calculated results are shown in Fig.~\ref{fig:HERMESH} and Fig.~\ref{fig:COMPASSH}, together with the corresponding data. As can be seen, the magnitude of the asymmetry induced by our new distribution is of a few percent and comparable to the Cahn effect in various kinematical ranges. The new and Cahn asymmetry are both flavor blind and positive for $\pi^\pm$ inclusive production while Boer-Mulders effect is flavor dependent. Our asymmetry seems to be bigger than Cahn effect in the kinematic range of the COMPASS measurement, as displayed in Fig.~\ref{fig:COMPASSH}. Another feature is that our new asymmetry is more weakly dependent on the kinematical variables than Cahn effect. Especially, it does not vanish when $P_{hT}$ is approaching zero, as Cahn effect does. The new asymmetry only begins to drop to zero when $z$ is smaller than 0.2, as can be seen in Fig.~\ref{fig:HERMESH}. We arrive at the analogous conclusion for the identified charged pions and kaons produced by SIDIS off nucleon, hydrogen and deuterium targets.

It is concluded that the $P_{hT}$ behavior of the COMPASS data seems to be incompatible with the corresponding behavior of HERMES data by Barone $et. al$~\cite{Barone:2015ksa}. Our new distribution in Eq.~(\ref{eq:Cahnlikekts}) and Eq.~(\ref{eq:Cahnlikekta}) as functions of kinematical variables is unlike Cahn effect, which is clearly illustrated in Fig.~\ref{fig:HERMESH} and Fig.~\ref{fig:COMPASSH}. So it would be helpful for the understanding this inconsistency phenomenologically. A reliable extract of Cahn and Boer-Mulders effects from experimental data should be done by considering the effect from our distribution ${\mathcal C}^{q}(x,\vec{r}_{\bot})$.

\section{Conclusion} \label{sec:conclusion}

In summary, we investigate the role of Pauli coupling in photon-quark vertex to SIDIS. In the calculation we utilize the scalar and axial-vector diquark models for the target nucleon. Our analytical formalism unveil that obvious $x$-dependence is introduced into PDFs by this additional photon-quark vertex. We determine the Pauli coupling in the scale of $Q^2_0$ = 0.26 GeV$^2$ by fitting to the global fit of the PDFs. The determined value of Pauli coupling agrees approximately with the prediction from instanton model. The numerical results have shown that Pauli coupling contributes considerably to the unpolarized and polarized PDFs in the low-$Q^2$.

Furthermore, we demonstrate that the Pauli coupling in photon-quark vertex can cause the helicity flip of the struck quark, leading to a new positive $\cos 2\phi_h$ asymmetry in SIDIS. The magnitude of the asymmetry is proportional to the squared Pauli coupling, which introduce a significant $Q^2$ dependence into its evolution. If admitting this $Q^2$ dependence expediently from instanton model, the given asymmetry is in the same level with Cahn effect and of the order of a few percent in size within the HERMES and COMPASS kinematics. In other words, the magnitude of this new asymmetry is expected to be larger than that of Cahn effect in the range of $Q^2 <$ 2.0 GeV since it has a stronger $Q^2$ dependence than Cahn effect. So the measurement of $Q^2$ dependence of the $\cos 2\phi_h$ asymmetry is crucial to disentangle higher-twist effects and our new asymmetry.

The important observation is that the $\cos 2\phi_h$ asymmetry at low $Q^2$ range should be explored systematically by including the effect from our distribution, together with Cahn and Boer-Mulders effects. The available data of $\cos 2\phi_h$ asymmetry in SIDIS do not allow yet a full extraction of Boer-Mulders function due to the present kinematics which is still dominated by the low-$Q^2$ region. It is already found that the higher-twist contributions are of great significance and strongly affect the results of the fits~\cite{Barone:2008tn,Barone:2009hw}. The multidimensional data for the multiplicities released by HERMES~\cite{Airapetian:2012yg} and COMPASS~\cite{Adolph:2014pwc} collaborations have big uncertainties and they are also not sufficient for a full determination of Boer-Mulders function~\cite{Barone:2015ksa}. Because contribution of our new asymmetry is sizable, it is more difficult for the experimental scrutiny of Boer-Mulders function. The future electron-ion colliders are anticipated to measure the $\cos 2\phi_h$ asymmetry in higher $Q^2$ range where both our new asymmetry and other high twist contribution, e.g. Cahn effect, are expected to be small. We do not find any contribution of Pauli coupling to $\cos \phi_h$ azimuthal modulation, so the condition would be more simple there for the determination of the Boer-Mulders function.

It has been discovered that Pauli couplings in both photon-quark and photon-gluon vertices are contributing to the angular dependence of SSA observables in polarized SIDIS~\cite{Cao:2017bdi,Hoyer:2005ev}. Our calculations here complement earlier studies and accomplish a full picture about the influence of the helicity nonconservation interactions on SIDIS measurements. The instantons are a possible underlying mechanism for these novel interactions and they give explicit $Q^2$ dependence of the corresponding couplings. Other optional approaches~\cite{Lu:2006kt,Bacchetta:2011gx}, e.g. Dyson-Schwinger equations with the nonperturbative
quark and gluon propagators~\cite{Chang:2010hb}, are waiting for extending to nonzero photon virtuality before a feasible comparison of different models.


\begin{acknowledgments}

It is gratefully acknowledged for the enlightening discussions with Prof. Vicente Vento, Prof. Nikolai Kochelev and Dr. Nikolai Korchagin at the initial stage of this work. We thank Dr. Ruilin Zhu and Dr. Wenjuan Mao for help to polish the paper. We would like to thank Prof. R.~Sassot for send us the files of DSS fragmentation functions, and Dr. Rong Wang for the IMParton PDF files. This work was supported by the National Natural Science Foundation of China (Grant No. 11405222).

\end{acknowledgments}

\appendix*

\section{$\vec{r}_{\bot}$-integrated distributions} \label{apdx:rt}

Her we list the $\vec{r}_{\bot}$-integrated results $f_1^{s,a} (x)$, $g_{1L}^{s,a}(x)$ and ${\mathcal C}^{s,a}(x)$ obtained in the context of our spectator diquark model:
\bea \label{eq:f1s}
f_1^{s}(x) &=& \frac{(1-x)^3}{(2\pi)^2} \frac{1}{24\,B_R^6(\Lambda_s^2)} \times \nonumber \\ &&
\begin{aligned}
\lf[ \lf(\mcf_D^q - \frac{\mcf_P^q}{2m_q} D_Q \rg)^2 \,B_R^2(\Lambda_s^2) + 2\,\lf(\mcf_D^q + \frac{\mcf_P^q}{2m_q} D_R \rg)^2 \,D_R^2 \rg. \\ \lf. + \lf(\frac{\mcf_P^q}{2m_q}\rg)^2 \,B_R^2(\Lambda_s^2) \,(2\,B_R^2(\Lambda_s^2)+D_R^2) \rg] \label{eq:f1a}
\end{aligned}
\\
f_1^{a}(x) &=& \frac{1-x}{(2\pi)^2} \frac{1}{24\,B_R^6(\Lambda_a^2)}\times \nonumber \\ &&
\begin{aligned}
\lf\{ 2\,(1-x)^2 \lf[ \lf( \mcf_D^q + \frac{\mcf_P^q}{2m_q} D_R \rg) D_R + \frac{\mcf_P^q}{2m_q} \frac{x}{1-x} \,B_R^2(\Lambda_s^2) \rg]^2 + \rg. \\ \lf. B_R^2(\Lambda_s^2) \lf[ \lf(\mcf_D^q + \mcf_P^q \rg)^2\,x^2 + \lf( \mcf_D^q  + \frac{\mcf_P^q}{2m_q} \lf( x\,D_R - D_Q \rg) \rg)^2 \rg. \rg. \\ \lf.\lf. + \lf(\frac{\mcf_P^q}{2m_q} \rg)^2 \,\lf(2\,(1+2\,x^2)\,B_R^2(\Lambda_s^2) + (1-x)^2\,D_R^2\rg) \rg.\rg. \\ \lf.\lf. - 2\, \lf( \mcf_D^q + \frac{\mcf_P^q}{2m_q} D_R \rg) \frac{\mcf_P^q}{2m_q} x(1-x) D_R \rg] \rg\} \qquad
\end{aligned}
\\ \label{eq:g1s}
g_{1L}^{s}(x) &=& \frac{(1-x)^3}{(2\pi)^2} \frac{1}{24\,B_R^6(\Lambda_s^2)}\times \nonumber \\ &&
\begin{aligned}
\lf[ - \lf(\mcf_D^q - \frac{\mcf_P^q}{2m_q} D_Q \rg)^2 \,B_R^2(\Lambda_s^2) + 2\,\lf(\mcf_D^q + \frac{\mcf_P^q}{2m_q} D_R \rg)^2 \,D_R^2 \rg. \\ \lf. + \lf(\frac{\mcf_P^q}{2m_q}\rg)^2 \,B_R^2(\Lambda_s^2) \,(-2\,B_R^2(\Lambda_s^2)+D_R^2) \rg]
\end{aligned}
\\ \label{eq:g1a}
g_{1L}^{a}(x) &=& \frac{1-x}{(2\pi)^2} \frac{1}{24\,B_R^6(\Lambda_a^2)}\times \nonumber \\ &&
\begin{aligned}
\lf\{ - 2\,(1-x)^2 \lf[ \lf( \mcf_D^q + \frac{\mcf_P^q}{2m_q} D_R \rg) D_R + \frac{\mcf_P^q}{2m_q} \frac{x}{1-x} \,B_R^2(\Lambda_s^2) \rg]^2 + \rg. \\ \lf. B_R^2(\Lambda_s^2) \lf[ \lf(\mcf_D^q + \mcf_P^q \rg)^2\,x^2 + \lf( \mcf_D^q  + \frac{\mcf_P^q}{2m_q} \lf( x\,D_R - D_Q \rg) \rg)^2 \rg.\rg. \\ \lf.\lf. + \lf(\frac{\mcf_P^q}{2m_q} \rg)^2 \,\lf(2\,B_R^2(\Lambda_s^2) - (1-x)^2\,D_R^2\rg) \rg.\rg. \\ \lf.\lf. + 2\, \lf( \mcf_D^q + \frac{\mcf_P^q}{2m_q} D_R \rg) \frac{\mcf_P^q}{2m_q} x(1-x) D_R \rg] \rg\} \qquad
\end{aligned}
\eea
\bea \label{eq:Cahnlikes}
{\mathcal C}^s(x) &=& \frac{(1-x)^3}{(2\pi)^2} \frac{1}{24\,B_R^4(\Lambda_s^2)} \lf(\frac{\mcf_P^q}{2m_q}\rg)^2\,2 x M D_R
\\ \label{eq:Cahnlikea}
{\mathcal C}^a(x) &=& \frac{(1-x)^2}{(2\pi)^2} \frac{1}{24\,B_R^4(\Lambda_a^2)} \lf(\frac{\mcf_P^q}{2m_q}\rg)^2 \, \lf[ (1-x)D_R^2 +D_R D_Q + \frac{2\,B_R^2(\Lambda_s^2)}{1-x} x(1+x) \rg]
\eea

\end{document}